\documentclass[11pt]{article}

\usepackage[english]{babel}
\usepackage{amsmath,amssymb,amsthm,amsfonts}
\usepackage{graphicx}
\usepackage[sort&compress,numbers,super]{natbib}
\usepackage{doi}
\usepackage{hyperref}

\usepackage{geometry}
\newcommand{\eps}{\varepsilon}

\newcommand{\mc}[1]{\mathcal{#1}}

\renewcommand{\H}{\mc{H}}

\newcommand{\bk}{\textbf{k}}

\renewcommand{\dag}{\dagger}
\newcommand{\nodag}{{\vphantom\dag}}

\newcommand{\cn}{c^{\nodag}}
\newcommand{\cd}{c^\dag}

\newcommand{\up}{\uparrow}
\newcommand{\down}{\downarrow}

\newcommand{\s}{\sigma}
\newcommand{\m}{\varepsilon}

\newcommand{\myonlinecite}[1]{\hspace{-1 ex} \nocite{#1}[\hspace{-1 ex} \citenum{#1}]}

\begin{document}

\title{Phase coexistence in a half-filled extended Hubbard model}
\author{Lo\"ic~Philoxene, Vu Hung~Dao and Raymond~Fr\'esard}
\date{
{\it Normandie Universit\'e, ENSICAEN, UNICAEN, CNRS, CRISMAT, 14000 Caen, France}\\[2ex]
\today}

\maketitle

\begin{abstract}
In this Letter we analyze the coexisting ordered phases that arise in the 
half-filled $\varepsilon$-$t$-$U$-$V$ extended Hubbard Model on the square lattice 
when tackled within the Kotliar and Ruckenstein slave boson representation in 
the thermodynamical limit. Particular emphasis is put on the dependence of the 
quasiparticle dispersions, the gaps, the spin-resolved renormalization factors, 
and the distribution of double occupancy on the microscopical parameters of 
the model. Our calculations are performed in a parameter range that is most 
suitable for resistive switching. In particular, the main contributions to the 
gaps are shown to either arise from $\varepsilon$ in the weak coupling regime, 
or from $U$ and $V$ when they are the largest scales. The gaps are thus from 
joint spin and charge origin, and so are the order parameters.

\end{abstract}

\section{Introduction}

Stripe order made of magnetic moments and charges belongs to the hallmarks of 
doped layered Mott insulators. The modulated electronic phases consist of 
two-dimensional antiferromagnetic domains separated by hole-rich linear walls.  
Since their discovery in the superconducting Sr-doped La$_2$CuO$_4$ cuprates 
(LSCO)~\cite{tranquada1995,tranquada2020}, they keep on being the subject of 
intense investigations~\cite{wu2021,campi2022} .  
It would be fair to note, however, that they had already attracted a significant
interest when they were first reported in doped nickelates~\cite{tranquada1994,
sachan1995}. Furthermore, they have been evidenced in other families of 
transition-metal oxides as well, such as layered cobaltates~\cite{cwik2009} and 
layered manganites~\cite{ulbrich2012}. 
Yet, not all stripy materials are alike. Indeed, the modulations are coined as 
diagonal or vertical, depending on the direction taken by their wave-vector in 
the Brillouin zone. Diagonal stripes are observed in all of these families of 
compounds, where they are found to be insulating~\cite{cheong1994,huecker2007}. 
In contrast, vertical modulations are reported in cuprates, only, and they are 
associated with a metallic or, even, a superconducting state~\cite{huecker2011}. 
For the latter group of oxides, model calculations have been able to establish 
a systematics of insulating diagonal filled and metallic half-filled stripes, 
with nearly one hole in the domain wall in the former case, and about half a 
hole in the latter case~\cite{raczkowski2006_b}. 
But the interplay with d-wave superconductivity is more challenging to unravel. 
As a solution of the two-dimensional Hubbard model, stripe order has been found 
to be very close or lower in  energy to its superconducting counterpart within 
a variety of numerical approaches \cite{himeda2002,ido2018,white2009,
huang2018,corboz2014,ponsioen2019,li2021}. Coexistence of both phases has 
been predicted, too \cite{leprevost2015,jiang2020}, but not necessarily in the 
ground state \cite{zheng2017}. 

The origin of the stripe order can be qualitatively understood as follows 
for the Hubbard model on a square lattice with on-site Coulomb repulsion. 
At half filling, the ground state is the antiferromagnetic N\'eel state. 
Hole doping induces an electron deficiency which results into phase separation 
on the nanoscale. It is accommodated by locating the added holes in a 
periodic structure of microscopic lines. This spatial distribution minimizes 
the energy as compared to an homogeneous solution. 
The separation between hole stripes thus increases with decreasing doping 
until it diverges in the limit of half filling, and the charge modulation 
vanishes. One may then presume that another mechanism is necessary in order 
to stabilize a spin and charge modulation at exactly zero doping. However, 
finding such a phase in extensions of the model turned out 
unsuccessful~\cite{vanDongen1994a,vanDongen1994b,wolff1983,deeg1993,
hirsch1984,segpunta2002,zhang1989,terletska2017,paki2019}. Hence, in an 
earlier study~\cite{philoxene2022}, we introduced in this perspective a 
Hubbard Hamiltonian extended with a nearest-neighbor interaction term of 
amplitude $V$, as well as a checkerboard modulation $\pm \varepsilon$ of 
the energy levels which straightforwardly induces charge order in the 
antiferromagnetic phase. 
We then found that the  model harbors charge order (CO) 
and joint spin and charge modulations (SCO) at finite values of the 
potential $\varepsilon$. The CO phase is realized by a staggering of
compensating charge excess or charge depletion in both directions, and 
the  SCO phase is realized by a staggering of compensated magnetic
moments accompanying the above CO. The latter are reminiscent of 
stripes~\cite{wu2021,campi2022,tranquada1994,sachan1995,tranquada1995,
cwik2009,ulbrich2012,cheong1994,huecker2007,huecker2011,yamamoto2007,
schwingen2008,schwingen2009,raczkowski2006_a,raczkowski2006_b,himeda2002,
ido2018} (for a review, see, \textit{e.g.}, 
Ref.~\myonlinecite{tranquada2020}), yet in the absence of hole doping. 

The aim of the present paper is to deepen our understanding of the phases 
that have been unveiled in our previous paper, and that could be suitable 
for resistive switching. Indeed, we have determined that CO and SCO 
can coexist in parameter domains at the boundary between their respective 
regions of stability. The zero-temperature transition between the 
two phases is found to be first order, which could be harnessed for 
application in digital electronics. Indeed, resistive switching in Mott 
insulators currently drives a considerable research activity since the 
property could enable a variety of novel functions, such as energy-efficient 
data processing for artificial intelligence~\cite{delvalle2018,delvalle2019,
bauers2021}, resistive RAM for data storage~\cite{janod2015,wang2019}, or 
optoelectronics~\cite{liao2018,coll2019}.
Furthermore, functional oxides also stand as promising replacements for 
conventional semiconductor materials when the latter will reach their 
intrinsic limitations at the nanoscale~\cite{takagi2010,coll2019}. 
V$_2$O$_3$ provides a prominent example~\cite{chudnovskii1998,brockman2014} 
but others are found among transition metal oxides such as 
VO$_2$~\cite{driscoll2009,kim2010}, NbO$_2$~\cite{liu2011,kumar2017}, and 
Ca$_2$RuO$_4$~\cite{nakamura2013,pan2014}.

This paper is organized as follows: In Sec.~\ref{model}, the extended Hubbard
model is presented, and the quasi-particle dispersion resulting from its 
treatment in the Kotliar and Ruckenstein slave-boson formalism is given.
In Sec.~\ref{results}, we perform a comparison of the band structure,
of the charge order parameters, as well as of the spin-resolved renormalization
factors between the charge ordered phase and the spin and charge ordered
phase. We focus on the intermediate and stronger coupling regimes where resistive
switching is most likely.
Lastly, Sec.~\ref{conclu} concludes on these results and gives a brief outlook.

\section{Model and methods}\label{model}

\subsection{Extended Hubbard model}

In our earlier study~\cite{philoxene2022}, we extended the Hubbard model in 
two ways. First, an efficient 
screening of the long-range Coulomb interaction is assumed, which reduces the 
latter to local and nearest-neighbor couplings, only. Second, a staggered crystal 
field that may arise from specific environments of the ions on which the interest 
is focused, is taken into account. For instance, in the context of LSCO 
superconductors, alternating La and, \textit{e.g.}, Pr chains, could produce 
such a contribution to the crystal field, as the electronic cloud of the La cations
is more extended than the one of the Pr cations. This results in a staggered 
$2\varepsilon$ energy distribution between adjacent lattice sites. One then 
arrives at the model Hamiltonian
\begin{equation}
{\cal H} = {\cal H}_0  + {\cal H}_{\varepsilon} + {\cal H}_U + {\cal H}_V. 
\end{equation}
Here the one-body part ${\cal H}_0  + {\cal H}_{\varepsilon}$, in the 
grand canonical ensemble,  is made of
\begin{equation}
\H_0  = \sum_{i,j,\s} t^{\phantom{0}}_{ij} \cd_{i,\s} 
\cn_{j,\s} -\mu \sum_{i,\s} \cd_{i,\s} \cn_{i,\s} ,
\end{equation}
where $\cd_{i,\s}$ ($\cn_{i,\s}$) creates (annihilates) an electron with spin 
$\s \in \{ \up,\down \}$ at the lattice site $i$, $t^{\phantom{0}}_{ij}$ is the 
hopping amplitude between lattice sites $i$ and $j$, and $\mu$ is the chemical 
potential controlling the band filling. This is supplemented by a staggered local 
potential of the form 
\begin{equation}
\H_{\m} = \sum_{i,\s} \m_i \cd_{i,\s} \cn_{i,\s} ,
\end{equation}
with
\begin{subequations}
\begin{align}
&\m_i = -\m \quad\text{if}\quad  i \in A , \\
&\m_i = +\m \quad\text{if}\quad  i \in B ,
\end{align}
\end{subequations}
where $A$ and $B$ are the two sublattices of a bipartite lattice. 
The next contribution is the local interaction term
\begin{equation}
\H_U = U \sum_i \cd_{i,\up} \cd_{i,\down} \cn_{i,\down} \cn_{i,\up} ,
\end{equation}
where  $U$ is the on-site Coulomb interaction strength.
The last term models the intersite Coulomb interaction as
\begin{equation}\label{model:HV}
\H_V = \frac{1}{2} \sum_{\substack{ \langle i,j \rangle \\ \s,\s'}} V_{ij} \cd_{i,\s} \cd_{j,\s'} \cn_{j,\s'} \cn_{i,\s} ,
\end{equation}
where $V_{ij}$ is the coupling restricted to nearest-neighbor sites $i$ and $j$. The $1/2$
factor accounts for the double-counting of the $ \langle i,j \rangle \equiv 
\langle j,i \rangle$ pairs in the sum.

Throughout the present study, we consider hopping to nearest neighbors, only, and 
we work in the half-filled subspace of the square lattice, such that
\begin{equation}
\sum_{i,\sigma} \langle n_{i,\sigma} \rangle=  N_L ,
\end{equation}
where $n_{i,\s}$ is the usual electron number operator ($n_i=n_{i,\up}+n_{i,\down}$), 
and $N_L$ is the total number of lattice sites. We also 
set the lattice parameter $a=1$ thereby fixing the length scale.

\subsection{The Kotliar and Ruckenstein representation}

We treat the Hamiltonian in the Kotliar and Ruckenstein 
representation~\cite{kotliar1986} on the saddle-point level of approximation. It is a well
established approach, which has been tested in many ways~\cite{lilly1990,fresard1991,
fresard1995,zimmermann1997,dao2017,riegler2020,fresard1992b,steffen2017} (for key 
fundamental aspects see 
Refs.~[$\!\!\!$\nocite{fresard1992,florens2002,fresard2001,fresard2007,
fresard2012b,dao2020}\citenum{fresard1992,florens2002,fresard2001,fresard2007,
fresard2012b,dao2020}]), and applied to a series of problems in the thermodynamical
limit~\cite{fresard1995,igoshev2015,igoshev2021,lilly1990,fresard1991,riegler2020,
raczkowski2006_c,seibold1998,zimmermann1997,fresard2002,fresard2022}. 
Here, we closely follow Ref.~\myonlinecite{philoxene2022}, and we stick to the notations 
and conventions used therein, up to one non-essential difference: instead of 
accounting for the staggered potential in the bosonic sector, we here do it in the 
fermionic one. Both calculations yield the same energy and boson expectation values,
but the quasi-particle dispersion is modified into
\begin{equation}\label{eq:disp_sb}
E_{\bk,\s,\nu} = \overline{\beta}_\s - \mu 
+ \nu \sqrt{ ( \tilde{z}^2_{\s} t^{\vphantom{0}}_{\bk} )^2 + (\Delta^{SB}_{\s} - \varepsilon)^2} ,
\quad \nu=\pm1 
\end{equation}
which reduces to the bare dispersion in the non-interacting limit, in contrast to Eq.~(31) in
Ref.~\myonlinecite{philoxene2022}.

\section{Results}\label{results}

\begin{table}
\centering
\begin{tabular}{c c c c}
Parameter set & $U/t$ & $V/t$ & $\eps/t$ \\
\hline\hline
I   & $4$ & $0.5$ & $0.1$ \\
II  & $4$ & $1.2$ & $0.1$ \\
III & $8$ & $1.8$ & $0.1$ \\
IV & $8$ & $2.5$ & $0.1$ \\
\hline\hline
\end{tabular}
\caption{Parameter sets used in Section~\ref{results}.}
\label{table}
\end{table}

Following Ref.~\myonlinecite{philoxene2022}, we shall mainly present results for 
two representative values of $\eps$ in this study, namely $\eps = 
0.1~t$ and $\eps = t$, as the proposed nature of the staggered
local potential does not suggest large values for $\eps$.
Moreover, we focus on the case of a half-filled band.
In our earlier work, one of the main results was that, for finite $\eps$,
the CO and SCO phases coexist in large regions of the $(U,V)$ 
parameter space, but little detail was provided regarding the nature
of this coexistence.
Evidence was also given that the properties of the phases  
qualitatively differ in the intermediate coupling regime from 
the strong coupling regime.

In the limit $t\rightarrow 0$, the study of our model considerably 
simplifies and we are left with two potential ground states: (i) a 
pair density wave of energy per site $U/2-\eps$, and (ii) an homogeneous 
state with one electron per site, of energy per site $2V$. Both are degenerate
for $U=4V+2\eps$ and one expects a finite hopping to modify this phase
boundary, and even to transform the associated phase transition into
a continuous one. Yet, according to Ref.~\myonlinecite{philoxene2022}, this does 
not happen, but the phase transition remains close to the $U=4V+2\eps$
line. Also, a great deal of information arises from the quasiparticle
dispersion and order parameters, on which this study focuses.
They are presented in this section, for four representative
parameter sets, given in Table~\ref{table}.

\subsection{Intermediate coupling}

According to the above, we focus the first part of our study on the 
intermediate coupling value of $U=4t$, and address the parameter sets
I and II, which are located on both sides of the $U=4V+2\eps$ line.
In this section, we thus present our results concerning the 
properties of both phases for these two parameter sets.

\begin{figure}
\centering
\includegraphics[width=1.\textwidth]{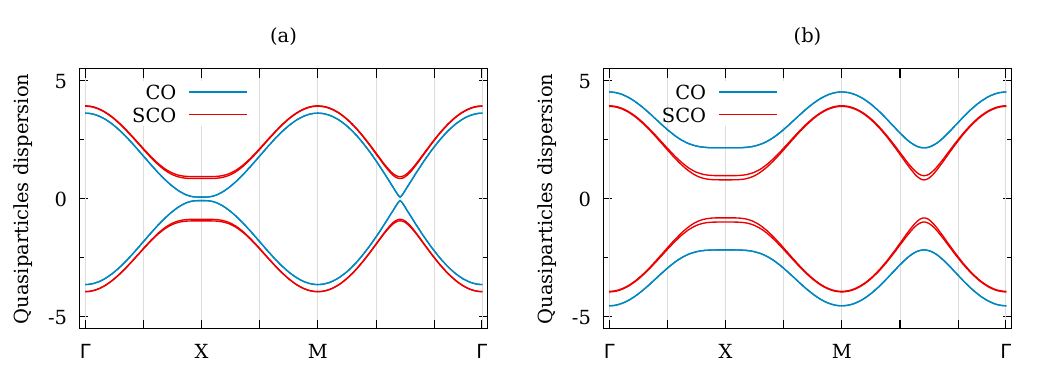}
\caption{
Band structure of the CO and SCO phases at $U=4~t$, $\eps=0.1~t$ 
and $V=0.5~t$ (a), and $V=1.2~t$ (b).
}
\label{bands_int}
\end{figure}

Fig.~\ref{bands_int}(a) displays the quasiparticle dispersion in both 
phases for the parameter set I.
In this case, the SCO phase is the ground state while the CO phase 
is metastable.
We notice that the CO phase is only weakly gapped, with a gap of 
about $0.14~t$. This, as argued below, hints at the existence of a 
parameter range, surrounding the present one, in which pure charge 
ordered solutions displaying modest charge modulations may be found.
In this case, the couplings are too weak to induce sizable charge
modulations in the phase. 
As for the SCO phase, it remains gapped in this region, 
with a gap of about $1.87~t$/$1.72~t$ for each spin branch, respectively.
For these parameter values, the
antiferromagnetic ordering of the SCO phase dominates, which leads 
to a band structure qualitatively similar to that of the pure spin-density
waves of the $t$-$U$-$V$ model. We begin to discern, however, the 
peculiar four-band structure characteristic of the SCO phase, but 
the fact that these bands remain pairwise nearly degenerate implies 
that, aside from an expected modest charge modulation, the charge 
distribution should still appear as predominantly 
homogeneous for these parameter values.
Besides, we find the two independent subbands of the SCO phase to be 
$3.00~t$/$3.06~t$ wide, while the width of the subband of the CO 
phase is $3.55~t$. They are hence narrower in the former case, as 
expected from the larger gap.

In Fig.~\ref{bands_int}(b), we show the impact of an increase in $V$
on the band structure of both phases. We take $V=1.2~t$ as a 
representative value, that is, for the parameter set II. 
In this case, the ground state is now the CO phase, 
with the SCO phase being metastable.
We further see that the gap open in the CO phase considerably increased
from $0.14~t$ to about $5~t$, which suggests a more firmly established 
charge order.
Besides, the quasiparticle dispersion of the SCO phase did not qualitatively 
evolve from the previous case, the only noticeable modification 
being that the difference in the respective gaps of both 
spin-projection branches of the dispersion has now increased to 
$1.61~t$ and $1.96~t$.
This could be interpreted as the influence of a more pronounced 
charge order in the SCO phase, due to the larger value of $V$, which
increasingly lifts the degeneracy between the bands. The subband 
width also changed from $3.00~t$/$3.06~t$ to $2.96~t$/$3.11~t$.

\begin{figure}
\centering
\includegraphics[width=1.\textwidth]{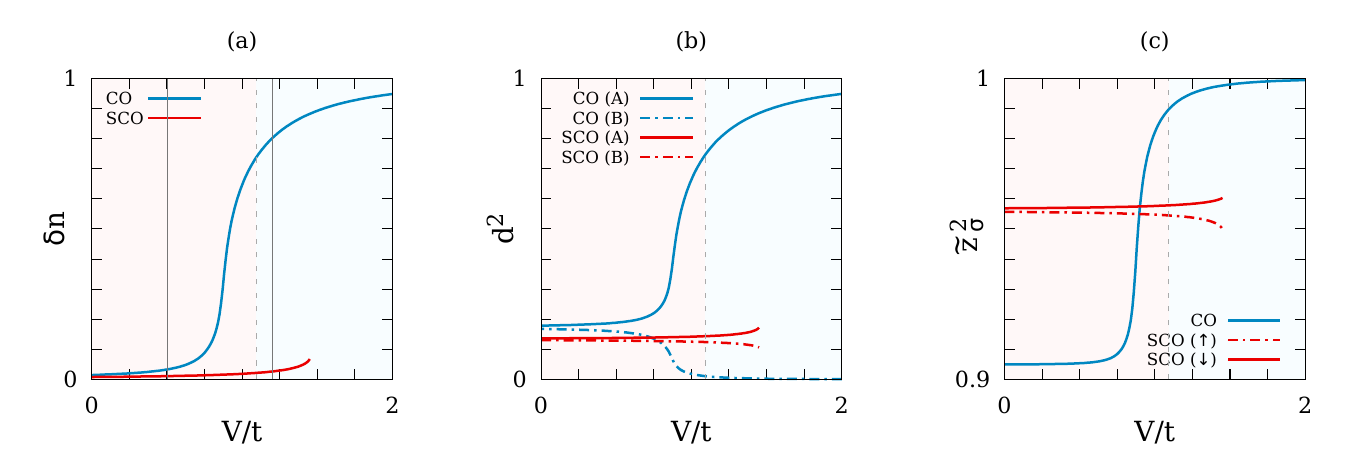}
\caption{
Charge modulation (a), sublattice resolved double occupancy (b), 
and spin resolved renormalization 
factor (c) of the CO and SCO phases for $U=4~t$ and $\eps=0.1~t$
as functions of $V$. The dashed gray line denotes the SCO-CO 
phase transition.
The solid gray lines denote the parameter sets I and II.
}
\label{op_int-0.1}
\end{figure}

The saddle-point values of the physical quantities of interest are 
shown in Fig.~\ref{op_int-0.1} as functions of $V$, for fixed 
$U=4~t$ and $\eps=0.1~t$. They involve the charge modulation 
\begin{equation}
\delta n = (n_A-n_B)/2,
\end{equation}
the sublattice resolved double occupancy 
\begin{equation}
d^2_{A/B} = \langle d^2 \rangle \pm \delta d^2,
\end{equation}
and the spin resolved renormalization factors 
\begin{equation}
\Tilde{z}^2_\sigma = z_{A,\sigma} z_{B,\sigma},
\end{equation}
where we introduced the sublattice and spin resolved Gutzwiller 
factors $z_{A/B,\sigma}$ (see, e.g., Eq.~(34b) in Ref.~\myonlinecite{philoxene2022}).
As can be seen from Fig.~\ref{op_int-0.1}(a) and Fig.~\ref{op_int-0.1}(b), 
for small values of the intersite coupling ($V \lesssim 0.5~t$),
the charge order is only weakly established, even in the CO phase, 
as the values of $\delta n$ and $\delta d^2$ are smaller than $10^{-2}$.
This owes to the fact that the relevant energy 
scale for such small values of $V$ is the on-site coupling $U$, 
favoring the antiferromagnetic arrangement of the spins with no 
charge modulation at half-filling. When increasing $V$, one notices
a slight increase in the charge ordering of the SCO phase, with
$\delta n$ eventually going from $7.39\times 10^{-3}$ for $V=0$ to 
$6.72\times 10^{-2}$ for $V=1.45~t$ ($\delta d^2$ takes similar values 
in this range of $V$). Beyond this value of $V$, the SCO solutions to 
the saddle-point equations (SPE) vanish and only CO solutions remain. 
We thus refer to the line in $(U,V)$ space at which the SCO solutions 
vanish as the SCO end-line. Conversely, the CO solutions remain stable 
for every shown values of $V$, leading to a nearly fully-established 
charge order for $V=2~t$, with values of $\delta n$ and $\delta d^2$
close to $0.95$. For this moderate $V$, yet fulfilling $V > U/4-\eps/2$, 
we thus meet a phase which strongly resembles a pair density wave, 
with one sublattice being nearly fully doubly-occupied, while the 
other sublattice remains essentially empty. 

These order parameters also help to qualitatively understand the phase
diagram. When the ground state is the SCO phase, for low values of 
$V$, the $-4V \delta n^2$ contribution to the free energy is 
irrelevant in both phases, and their respective free energies 
essentially differ via the term $U \langle d^2 \rangle$.
Since the CO phase has the largest value of $\langle d^2 \rangle$,
it naturally has a larger energy, leading to the SCO 
ground state. At larger values of $V$, however, the $-4V \delta n^2$
becomes predominant and, as $\delta n$ is an order of magnitude larger
in the CO phase than in the SCO one, the ground state thus becomes CO.

Regarding the renormalization factors, which are displayed in 
Fig.~\ref{op_int-0.1}(c), one can see that it is minimal in the CO 
phase for small values of $V$ --- with a value of about $0.95$.
It however harbors a sharp increase for values of $V$ 
ranging from $0.75~t$ to $t$, corresponding to the value at which 
the charge order begins to sharply increase in the CO phase as well.
This indicates that the charge modulations tend to hinder the 
correlations originating from the on-site interactions, as they are 
produced for smaller values of the ratio $U/V$. 
In the SCO phase, however, the renormalization factors for the up 
and down spin branches of the dispersion remain very close to one another 
for small values of $V$, with values averaging to $0.98$ and a relative 
difference of roughly $3\times 10^{-4}$ for $V=0$.
This relative difference slightly increases as $V$ is 
increased, but remains small as it only reaches approximately 
$3\times 10^{-3}$ at the SCO end point.

\begin{figure}
\centering
\includegraphics[width=1.\textwidth]{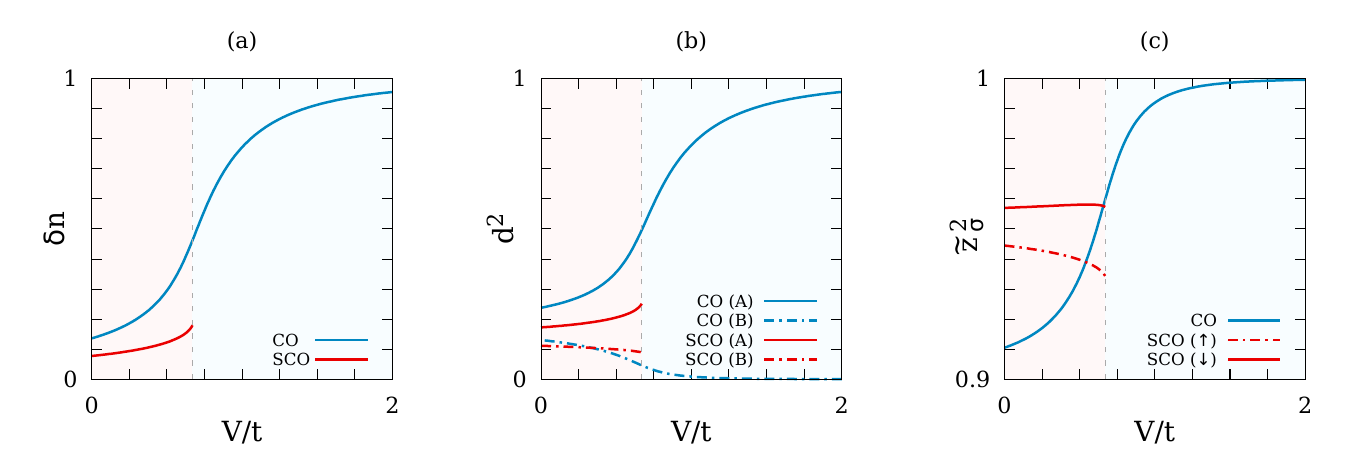}
\caption{
Charge modulation (a), sublattice resolved double occupancy (b), 
and spin resolved renormalization 
factor (c) of the CO and SCO phases for $U=4~t$ and $\eps=t$
as functions of $V$. The dashed gray line denotes the SCO-CO 
phase transition.
}
\label{op_int-1}
\end{figure}

In Fig.~\ref{op_int-1}, the same quantities as in 
Fig.~\ref{op_int-0.1} are shown, again as functions of $V$ and for 
fixed $U=4~t$ but we now increase the staggered potential to $\eps=t$.
We notice from Fig.~\ref{op_int-1}(a) and Fig.~\ref{op_int-1}(b) that 
the charge order is now more pronounced, even at $V=0$, confirming 
that larger values of $\eps$ tend to increase the charge modulations 
even when $U$ is the dominant energy scale. For these values of $U$ and 
$\eps$, the SCO-CO transition line corresponds to the SCO end-line, 
since in this regime, the SCO solutions to the SPEs vanish while still 
being lower in energy than the CO solutions. 
In the SCO phase, we again see that the charge modulations become 
increasingly pronounced as $V$ is increased towards the SCO 
end-point, with $\delta n$ ranging from $0.08$ to $0.18$ for $V$ 
between $0$ and $0.67~t$, and $\delta d^2$ ranging from $0.03$ to 
$0.08$ in the same $V$-interval. The stronger dependence of
these quantities on $V$ as compared to the $\eps=0.1~t$ case implies
that the effect of the intersite coupling is enhanced by the larger 
residual charge order caused by $\eps$ at $V=0$.
Besides, in the CO phase, the situation is qualitatively similar to 
the $\eps=0.1~t$ case for large values of $V$. The main difference 
lies in the $V \lesssim 0.75~t$ range, in which the minimal 
charge order imposed by $\eps=t$ causes the CO solutions to be more 
sensitive to variations in $V$.

Regarding the renormalization factors in Fig.~\ref{op_int-1}(c), we 
see that the relative difference between both spin branches of the 
dispersion in the SCO phase has increased, with $\Tilde{z}_\sigma 
\sim 0.975 \pm 0.003$ for $V=0$ and $\Tilde{z}_\sigma \sim 0.972 \pm
0.006$ for $V=0.67~t$. As argued above, this owes to the stronger 
charge order in the SCO phase at $\eps=t$, as compared to the $\eps=
0.1~t$ case. For the CO phase, the renormalization factor takes the 
value $0.91$ for $V=0$, indicating a slight increase in the effect 
of the local correlations as compared to the previous case. Similarly 
to its $\eps=0.1~t$ counterpart, the renormalization factor 
increases with $V$, up to $\Tilde{z}_\sigma=1$, but with a 
smaller slope in this case --- suggesting a smoother transition 
from the residual charge ordered regime to the $V$-driven one.

Comparison of the interplay of the SCO and CO phases between the 
two presented values of $\eps$ thus showed that, on one hand, the 
nature of the SCO and CO phases remain qualitatively independent on 
$\eps$ when it is smaller than the $U$ and $V$ couplings. On the 
other hand, the charge order at low values of the intersite coupling 
seems to be strongly driven by $\eps$ in both phases, with larger 
values of $\eps$ implying not only stronger charge modulations, but 
an enhanced $V$-dependence of the charge order as well.
Additionally, we emphasize that all these quantities are subject to 
a discontinuous variation at the SCO-CO transition, which occurs 
when the free energy of both phases become degenerate for 
$\eps=0.1~t$ and when the SCO solutions vanish for $\eps=t$. 
This thus confirms the first-order character of the phase transition
between these two phases.

\subsection{Stronger coupling}

In the stronger coupling regime where $U$ is larger than the bandwidth, 
here $U = 8~t$, the energy scales of the interactions $U$ and $V$ 
prevail over the amplitude of the staggered local potential. 
In this way, one would expect to recover, at least qualitatively, 
the behavior of the well studied $t$-$U$-$V$ model at half-filling 
\cite{zhang1989,terletska2017,paki2019}, in which a 
spin-density wave and a charge-density wave compete at zero 
temperature, with a phase transition at $V=U/4$. 
In our study, the CO phase would then be the analog of the charge-
density wave, albeit not spontaneously formed due to the finite 
value taken by $\eps$, and the SCO phase would be the analog of the 
spin-density wave favored by the local interaction $U$.

\begin{figure}
\centering
\includegraphics[width=1.\textwidth]{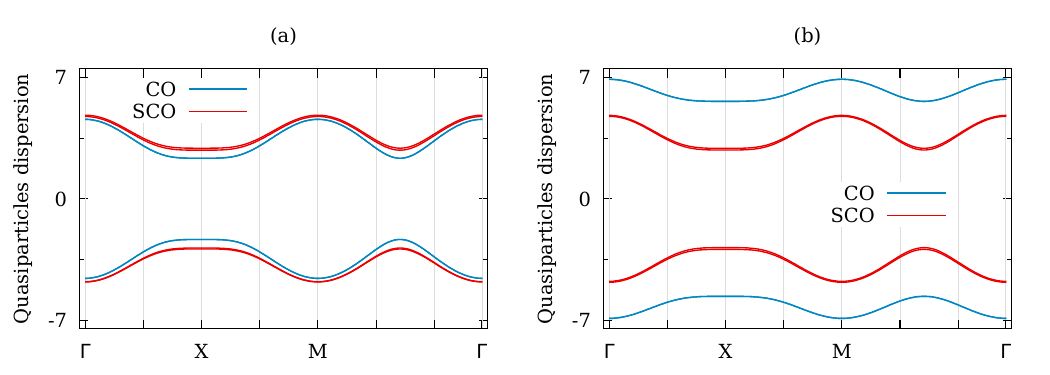}
\caption{
Band structure of the CO and SCO phases at $U=8~t$, $\eps=0.1~t$ 
and $V=1.8~t$ (a), and $V=2.5~t$ (b).
}
\label{bands_str}
\end{figure}         

Fig.~\ref{bands_str}(a) displays the band structure of both 
phases for the parameter set III.
In this case, the SCO phase is the ground state. One can see that 
both phases are strongly gapped in this regime, with a gap of about 
$4.6~t$ in the CO phase and $6~t$ in the SCO phase. Additionally, 
Fig.~\ref{bands_str}(b) presents the band structure of both phases 
for the parameter set IV, in which $V$ is increased to $2.5~t$. This 
corresponds to a region of the phase diagram in which the CO phase 
is the ground state, while the SCO phase is metastable. Compared to 
Fig.~\ref{bands_str}(a), the gap of the CO phase increased more than
twofold, reaching a value of about $11.2~t$. This implies that the
charge order is much more firmly established in the CO phase for $V=
2.5~t$ than it is for $V=1.8~t$. 
As for the SCO phase, one can see that its
band structure has barely evolved after the increase in $V$. 
Moreover, no noticeable lifting of the pairwise degeneracy of the bands
in this phase is observed.
This hints at the fact that, in this regime, the SCO phase remains 
qualitatively independent on the value of $V$ and is essentially a 
slightly deformed version of the pure spin-density wave of the 
$t$-$U$-$V$ model.

\begin{figure}[!h]
\centering
\includegraphics[width=1.\textwidth]{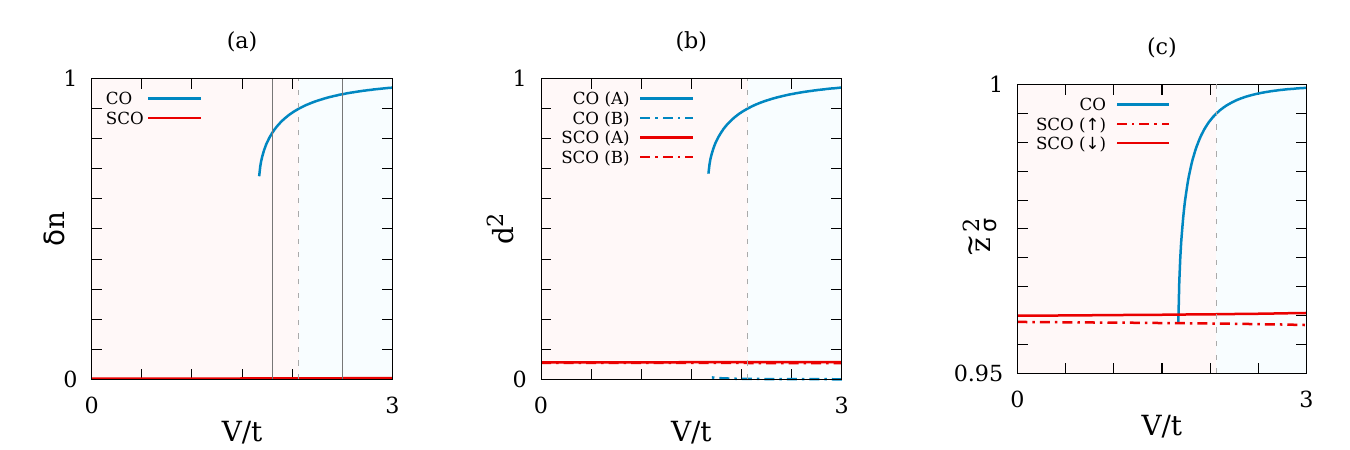}
\caption{
Charge modulation (a), sublattice resolved double occupancy (b), 
and spin resolved renormalization 
factor (c) of the CO and SCO phases for $U=8~t$ and $\eps=0.1~t$
as functions of $V$. The dashed gray line denotes the SCO-CO 
phase transition.
The solid gray lines denote the parameter sets III and IV.
}
\label{op_str-0.1}
\end{figure}

In order to verify these interpretations, let us now turn to the 
saddle-point values of the quantities of interest in this stronger 
coupling regime.
The charge modulation, the double occupancy and the renormalization 
factors for the parameter set $U=8~t$ and $\eps=0.1~t$ are displayed
in Fig.~\ref{op_str-0.1}, in dependence on $V$. One can see from 
Fig.~\ref{op_str-0.1}(a) and Fig.~\ref{op_str-0.1}(b) that the charge modulation in the 
SCO phase are barely noticeable, reaching, at most, $3.8\times 10^
{-3}$ for $V=3~t$. Likewise, the double occupancy is very weakly modulated 
as we find it to be $(5.56 \pm 0.17)\times 10^{-2}$ for $V=3~t$. This implies
that, in this regime, the charge distribution in the SCO phase is 
nearly homogeneous, and it is only the spin modulation that remains sizable
(see Fig.~4 in \cite{philoxene2022}).
As for the CO phase, we note the appearance of a CO end-line, below 
which the purely charge ordered solutions cease to exist. Above this
end-line, the charge modulation of the CO phase strongly depends on 
$V$, with $\delta n$ and $\delta d^2$ displaying a nearly vertical 
slope above $V=1.67~t$, at which $\delta n \sim 0.675$ and 
$d^2_{A,B} \sim 0.35 \pm 0.34$.
Then, at large $V$, $\delta n$ approaches $1$ while $d_{A,B}^2$ approaches
$0.5\pm0.5$, implying that the pair-density wave is almost fully
established.

One can infer, from the renormalization factors in 
Fig.~\ref{op_str-0.1}(c), that the situation is qualitatively similar
to that of the intermediate coupling case discussed above.
Indeed, the renormalization factors of the SCO phase show very weak
dependence on $V$, taking values of about $0.958$ for $V=0$ and 
increasing (or decreasing) by less than $3~\%$ when $V$ is increased
to $3~t$. Such a weak $V$-dependence of $\Tilde{z}_\sigma$ owes to 
the fact that, in this coupling regime, the SCO phase is largely 
dominated by its spin modulations, while the charges remain nearly 
homogeneously distributed, irrespective of the value of $V$.
In contrast, in the CO phase, the renormalization factor takes
the value $0.958$ at the end-point $V=1.67~t$, and sharply increases
with $V$, saturating to $1$ for large values of $V$. This again 
indicates that, in the large-$V$ CO phase, the static correlations 
induced by local interactions at half-filling are mostly suppressed
when a genuine pair-density wave sets in.

\begin{figure}
\centering
\includegraphics[width=1.\textwidth]{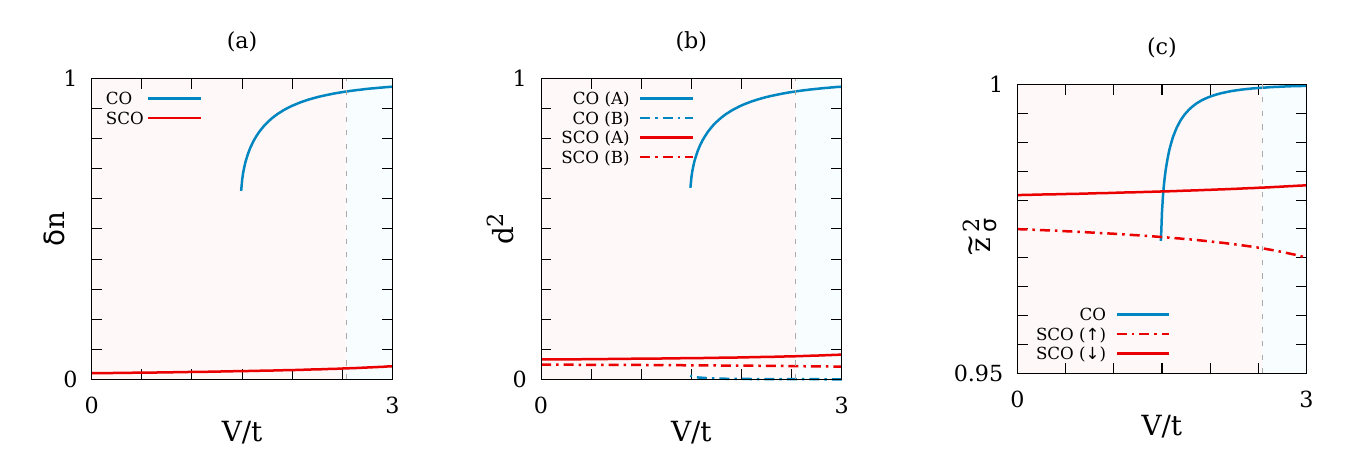}
\caption{
Charge modulation (a), sublattice resolved double occupancy (b), 
and spin resolved renormalization 
factor (c) of the CO and SCO phases for $U=8~t$ and $\eps=t$
as functions of $V$. The dashed gray line denotes the SCO-CO 
phase transition.
}
\label{op_str-1}
\end{figure}

In Fig.~\ref{op_str-1}, the same quantities are presented, as 
functions of $V$, for $U=8~t$ and for an increased $\eps=t$. As in 
the intermediate coupling case, we notice that the increase in $\eps$
induces an enhanced charge ordering in both phases. Specifically, 
one can see in Fig.~\ref{op_str-1}(a) and Fig.~\ref{op_str-1}(b) that the charge order 
in the SCO phase becomes noticeably $V$-dependent, with $\delta n$ 
ranging from $2.02\times 10^{-2}$ to $4.34\times 10^{-2}$, as well as 
$d_{A,B}^2$ ranging from $(5.7\pm0.8)\times 10^{-2}$ to 
$(6.2\pm2.0)\times 10^{-2}$, for $V$ between $0$ and $3~t$. In the CO phase, 
the situation is qualitatively similar to the $\eps=0.1~t$ case, 
with an end-point at $V=1.49~t$, below which no CO solutions to the 
SPE exist. At this end-point, we find $\delta n \sim 0.63$ and 
$d_{A,B}^2 \sim 0.32\pm0.31$, followed by a sharp increase towards the 
pair-density wave limit with $\delta n = 1$ and $d_{A,B}^2 = 0.5 \pm
0.5$ when increasing $V$. 

Fig.~\ref{op_str-1}(c) shows that, as in the intermediate coupling 
case, the increase in the staggered potential enhances the 
difference in the renormalization factors of the SCO phase, with 
$\Tilde{z}^2_\sigma \sim 0.956 \pm 0.006$ for $V=0$ and 
$\Tilde{z}^2_\sigma \sim 0.953 \pm 0.012$ for $V=3~t$. In terms of 
absolute values, these renormalization factors are a few percents 
lower than their $\eps=0.1~t$ counterpart. This is further evidence 
that the charge order induced by $\eps$ and $V$ in the SCO phase 
hinders the effect of local correlations, as, e.~g., for $\eps=V=0$ 
(pure spin-density wave) we find that $\Tilde{z}^2_\sigma$ takes the
value $0.905$ for $U=4~t$, and $0.619$ for $U=8~t$.
As for the CO phase, the tendencies are again the same as in the 
$\eps=0.1~t$ case, namely a sharp increase from the end-point
$V=1.49~t$, going from $\Tilde{z}^2_\sigma \sim 0.947$ to 
$\Tilde{z}^2_\sigma = 1$ at large $V$, when the pair-density wave 
becomes fully established.

In this stronger coupling regime, we have thus unraveled that the 
picture becomes simpler than at intermediate coupling. 
Indeed, on one hand, the SCO phase essentially reduces to its 
antiferromagnetic spin modulations, with small inhomogeneities in 
its charge distribution for sufficiently large values of $\eps$ and
$V$.  On the other hand, the CO phase, for values of the parameters 
for which it exists, strongly resembles a genuine pair-density wave, in
which one sublattice is totally filled while the remaining one is 
essentially empty.

Lastly, let us briefly discuss the strong coupling regime. When one 
of the energy scale becomes arbitrarily large, the picture simplifies
even more. Specifically, if $U$ is much larger than the bare bandwidth,
while $V$ and $\eps$ remain smaller than the latter, the SCO 
phase loses its dependence on $V$ and $\eps$, and a phase strongly resembling 
the genuine spin-density wave, with energy close to $2V$, is found.
As for the CO phase, it is simply not stabilized in this regime.
Conversely, if $V$ or $\eps$ become arbitrarily large as compared to
the bare bandwidth, while $U$ remains smaller than the latter, we 
find that no SCO solutions exist anymore, whereas the CO phase becomes
the genuine pair-density wave, with energy $U/2-\eps$.

\section{Discussion and summary}\label{conclu}

Summarizing, we presented above the quasiparticle dispersions and the
order parameters characterizing the two phases that coexist in the 
half-filled $\eps$-$t$-$U$-$V$ extended Hubbard model when tackled in the 
Kotliar and Ruckenstein slave-boson 
representation.
For $U \lesssim 4V+2\eps$, the ground state is charge ordered.
In that case, we obtain the renormalization factors $\tilde{z}_\sigma$
to modestly differ from one.
This also remains true in the regions of the phase diagram in which 
the ground state is spin-and-charge ordered.
Accordingly, we find it useful to use the Hartree-Fock approximation 
expressions of the gaps, since they are easier to interpret than their 
Kotliar and Ruckenstein slave-boson  counterpart. Following 
Ref.~\myonlinecite{philoxene2022}, they read
$\Delta^{HF}_\sigma = |(8V-U) \delta n + \sigma U m_z + 2 \eps|$,
with $\sigma=\pm 1$ and $m_z$ the staggered magnetization.
They highlight the complex origin of the gaps and their sophisticated
dependence on the parameters of the model.
In the non-interacting limit, the two gaps collapse into a single one
given by $2|\eps|$.
It therefore has a charge character, which unravels the physics embedded
in the crystal field.
This charge character of the gap is manifest in the limit of vanishing
$U$ and $\eps$, too, as well as when $U$ is the only vanishing parameter.
In this latter case, the gap never closes.
Indeed, positive $\eps$ pins the charge density wave to have its 
maxima on the A sites, rendering $\delta n$ positive.
Hence the gap remains open for $V>0$.
For $V<0$, with $4|V| \gg \eps$, the charge distribution remains 
largely homogeneous, so that $|4V\delta n|<\eps$ and no gap closing
takes place.
The case $\eps<0$ may be discussed in the very same fashion, after
having exchanged the A and B sublattices.
Conversely, when $U$ is the only non-vanishing parameter, both gaps 
merge into a single one, which now has a magnetic character.
Hence, at last, the two gaps that open in the SCO phase arise from
their joint charge and magnetic characters.

Quantitatively, the above Hartree-Fock expressions nicely explain 
the evolution of the  gaps found when performing the calculations 
with the parameter sets I--IV, which dictate the behavior of the 
above discussed subband  widths.
This follows from the finding that the spin-resolved renormalization
factors $\tilde{z}_\sigma$ remain
close to one in the entire phase coexistence region of the phase
diagram.

Also, it is worth noticing that the physics of joint spin and charge
modulations in the SCO phase is enabled by a finite, but not 
necessarily sizable, value of $\eps$.
The ensuing interplay between the CO and SCO phases is thus relevant
even when $\eps$ is orders of magnitude smaller than the energy
scale of the interactions, as is shown by the results for $\eps=0.1~t$
at intermediate and stronger coupling.

Lastly, let us emphasize that these large regions of phase
coexistence between two insulating phases harboring quantitatively 
distinct gaps shows great potential in the tailoring of materials
prone to resistive switching.

\section*{Acknowledgments}
This work was supported by R\'egion Normandie through the ECOH project.

\bibliographystyle{unsrt}

%

\end{document}